# STRUCTURAL FLYBY CHARACTERIZATION OF NANOPOROSITY

R. R. Rosa[*1], A. Ferreira da Silva[2], R. C. Brito[1], L. S. Roman[3], M. P. M. A. Baroni[1], F. M. Ramos[1], R. Ahuja[4] and C. Persson[5]

[1] Nucleous for Simulation and Analysis of Complex Systems, Lab for Computing and Applied Mathematics, National Institute for Space Research, 12245-970 S. J. dos Campos – SP - Brazil

[2] Instituto de Física, Universidade Federal da Bahia, 40210-340 Salvador – BA – Brazil.

[3] Instituto de Física, Universidade Federal do Paraná, 81531-990 Curitiba – PR – Brazil.

[4] Department of Physics, Uppsala University, SE-751 21 Uppsala - Sweden.

[5] Applied Materials Physics, Royal Institute of Technology, SE-10044 Stockholm - Sweden.

## Abstract

Recently, Ferreira da Silva et al. [3] have performed a gradient pattern analysis of a canonical sample set (CSS) of scanning force microscopy (SFM) images of π-Si. They applied the so-called Gradient Pattern Analysis to images of three typical π-Si samples distinguished by different absorption energy levels and aspect ratios. Taking into account the measures of spatial asymmetric fluctuations they interpreted the global porosity not only in terms of the amount of roughness, but rather in terms of the structural complexity (e.g., walls and fine structures as slots). This analysis has been adapted in order to operate in a OpenGL flyby environment (the StrFB code), whose application give the numerical characterization of the structure during the flyby real time. Using this analysis we compare the levels of asymmetric fragmentation of active porosity related to different materials as π-Si and "porous diamond-like" carbon. In summary we have shown that the gradient pattern analysis technique in a flyby environment is a reliable sensitive method to investigate, qualitatively and quantitatively, the complex morphology of active nanostructures.

# 1  Introduction

Most porous materials, such as porous silicon and porous diamond-like carbon, are composed by pores whose structural scales are no greater than the size of molecules (2-50nm). Usually, the porous silicon samples are produced by anodic etching of crystalline silicon (c-Si) wafers in hydrofluoric (HF) acid solution [3]. As reported by many authors (e.g.,[1,2]) one of the main problem in the phenomenology of porous silicon ($\pi$-Si) sample is that there is no satisfactory explanation on the possible correlation between their photoluminescence (PL) and their structural properties due to the different formation parameters (doping level, HF concentration and current density) [8]. In another hand, the morphological analysis of the "porous diamond-like" carbon structures, by field emission scanning electron microscopy, revealed that they had a highly porous structure, which was attributed to the modification of the kinetics of the carbon deposition process due to the presence of helium as a buffer gas [9].

Recently, Ferreira da Silva et al. [3] have performed a gradient pattern analysis of a canonical sample set (CSS) of scanning force microscopy (SFM) images of $\pi$-Si. They applied the so-called Gradient Pattern Analysis (GPA) [3-5] to images of three typical $\pi$-Si samples distinguished by different absorption energy levels and aspect ratios (low, intermediate and high roughness). Due the lack of robust tools for characterization of porous structures in nanometrics scales we will use in this work the GPA. The GPA is an innovative technique, which characterizes the formation and evolution of extended patterns based on the spatio-temporal correlations between large and small amplitude fluctuations of the structure represented as a gradient field [6].

Due to the high sensitivity of the asymmetric fragmentation parameter (the so-called first gradient moment $g_1^a$) to quantify asymmetric fine structures in complex extended patterns, a classification of the canonical $\pi$-Si samples, of the same size, using asymmetric fragmentation values, was used to characterize silicon porosity quantitatively. They showed that, for the canonical sample set, the only parameter showing a direct relationship of the structural asymmetry with PL energy was the first gradient moment ($g_1^a$). Taking into account this result they interpreted the global porosity not only in terms of the amount of roughness, but rather in terms of the structural complexity of the roughness, mainly that described by means of the shape and size of asymmetric main structures like walls and fine structures like slots. This analysis has been adapted in order to operate in a OpenGL flyby environment (the StrFB code), whose application give the numerical characterization of the structure during the flyby real time. It is hoped that the development of a classification methodology of porous materials can be important for nanofabrication technologies. In this paper, using this 3D computational analytical environment, we performed comparative structural analyzes of a π-Si sample and a porous diamond-like carbon structure as a preliminar example of a possible porous materials classification methodology.

## 2 The Gradient Pattern Analysis Formalism

The spatial structure fluctuation of a global pattern given by the matrix $M(x,y)$, can be characterized by its gradient vector field $G = \nabla[M(x,y)]$, which is composed by $V$ vectors $r$ where a vector $r_{i,j}$ is represented, besides its location $(i,j)$ in the lattice, by its norm $(r_{i,j})$ and phase $(\phi_{i,j})$, so that associated to each position in the lattice we have a respective vector $(r_{i,j} = (r_{i,j}, \phi_{i,j}))$. The local spatial fluctuations, between a pair of pixels, of the global pattern is characterized by its gradient vector at corresponding mesh-points in the two-dimensional space. In this representation, the relative values between pixels are relevant, rather than the pixels absolute values. Note that, in a gradient field such relative values, can be characterized by each local vector norm and its orientation. Thus, according to Rosa et al [6], a given matricial scalar field can be represented as a composition of four gradient moments: $g_1$, is the integral representation of the vectors distribution; $g_2$, is the integral representation of the norms; $g_3$, is the integral representation of the phases; and $g_4$, is the complex representation of the gradient pattern (Figure 1).

Considering the sets of local norms and phases as discrete compact groups, spatially distributed in a lattice, the gradient moments have the basic property of being, at least, rotational invariant. As we are interested in nonlinear extended structures we used a computational operator to estimate the gradient moment $g_1$ based on the asymmetries among the vectors of the gradient field of the scalar fluctuations. A global gradient asymmetry measurement, can be performed by means of the asymmetric amplitude fragmentation (AAF) operator[3]. This computational operator measures the symmetry breaking of a given dynamical pattern and has been used in many applications[3-6]. The measure of asymmetric spatial fragmentation $g_1^a$ is defined as

$$g_1^a \equiv (C - V_A) / V_A \mid C \geq V_A > 0 \qquad (1)$$

where $V_A$ is the number of asymmetric vectors and C is the number of correlation bars generated by a Delaunay triangulation having the middle point of the asymmetric vectors as vertices [6-7]. As an example, Fig.2 shows the contour pattern and the respectives gradient and triangulation fields for a sub-sample 8x8 of the sample showed in Figure 3a. Note that, Fig. 2 has just an illustrative character on the GPA operation in order to get the first gradient moment.

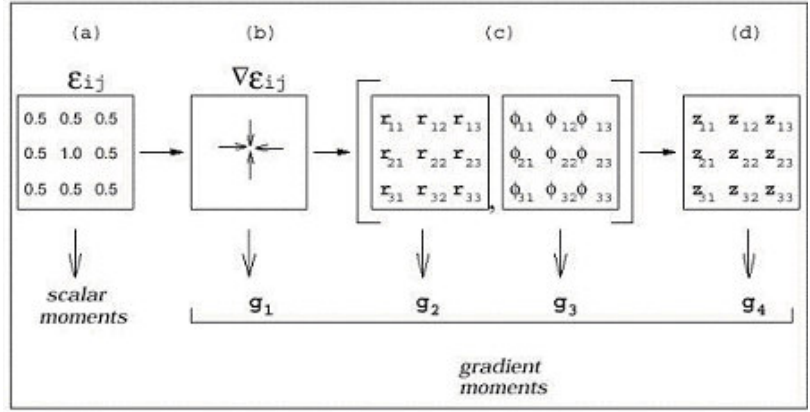

**Figure 1**: A schematic representation of the Gradient Pattern Analysis of a matricial scalar field: (a) an arbitrary normalized extended scalar field; (b) the corresponding gradient pattern of the amplitude fluctuations; (c) the norm and the phase of the fluctuations; (d) the complex representation of the fluctuations

The Delaunay triangulation $TD(C, V_A)$ is a fractional field with dimension less than two – the lattice dimension [4]. When there is no asymmetric correlation in the pattern, the total number of asymmetric vectors is zero, and then, $g_1^a$ is null. Otherwise, this parameter quantifies the level of asymmetric fluctuations [7].

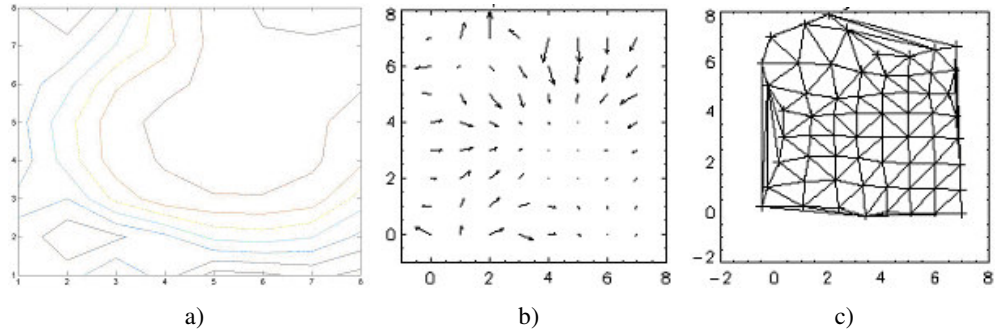

**Figure 2**: a) The amplitude contour of submatrix (8x8) of the porous silicon sample showed in Figure 3a; b) its respective gradient field representing the amplitude fluctuations and c) its respective triangulation field.

## 3   Data Analysis and Experimental Results

In this work, we have used two different typical-structure samples: the first is a sample of porous silicon characterized by intermediate roughness and the second is the sample of "porous diamond-like" carbon [9] (Figure 3 and 4). The measured parameters are $g_1^a$ (64), $g_1^a$ (8) and $s_a$. The parameters $g_1^a$ (64) and $g_1^a$ (8) are the amount of asymmetric fragmentation of a matrix of size 64x64 and of its sub-matrix of size 8x8. This sub-matrix is in center of a 64x64 matrix. The paramenter $s_a$ is the asymmetry scale defined as the difference between $g_1^a(64)$ and $g_1^a(8)$. The absolute (for 64x64 scale) and the average (for 8x8 scales) measures obtained from the application of the GPA Flyby are shown in Table 1.

**Table 1**: The measured parameters for each sample

| Sample | $g_1^a$(64x64) | $<g_1^a>$(8x8) | as |
|---|---|---|---|
| Porous Silicon | 1.99329 | 1.88889 | 0.10440 |
| Diamond | 1.98277 | 1.82812 | 0.15465 |

In this report we are not interested in measurements of the higher order gradient moments. Several calculations on random patterns have shown that $g_1$ gradient moment is much more sensitive and precise in characterizing asymmetric structures than the correlation length measures [3,4]. In order to calculate the gradient moments in a 3D dynamic environment we developed the StrFB code, a flyby "real time" analytical computational camera from where it is possible to visualize the shapes and sizes of local roughness related to the materials porosity. This computational environment for analytical visualization was developed using the graphic tool Open GL. The active window interface is able to show scales ranging from 3x3 up to 64x64 given online the respective values for the gradient moment $g_1$. A snapshot showing the asymmetric fragmentation of the amplitudes ($g_1^a$=1.88889) for a central region of the sample of porous silicon from our data set, is shown in Figure 3, and in Figure 4 is a snapshot of "porous diamond-like" carbon. A interesting preliminar result as that although the asymmetries found for the porous silicon are greater than that found for diamond-like carbon, the second one has a greater asymmetry scale. This kind of phenomenon we call "to prevail by scale asymmetry" (PSA).

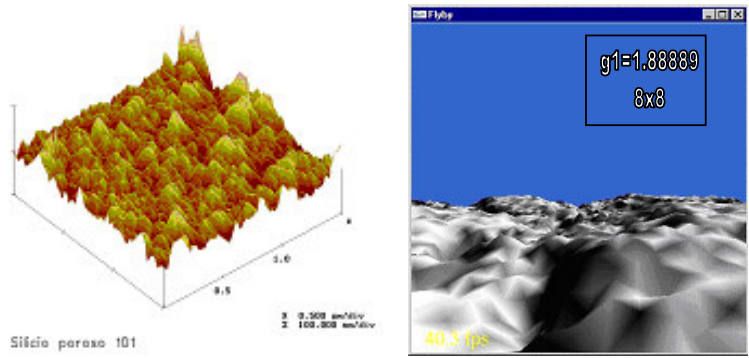

**Figure 3**: SFM structural pattern of a PSi sample (left) and and a flyby snapshot (right), using the StrFB code, computing the level of correspondent local asymmetry in an area of 8x8.

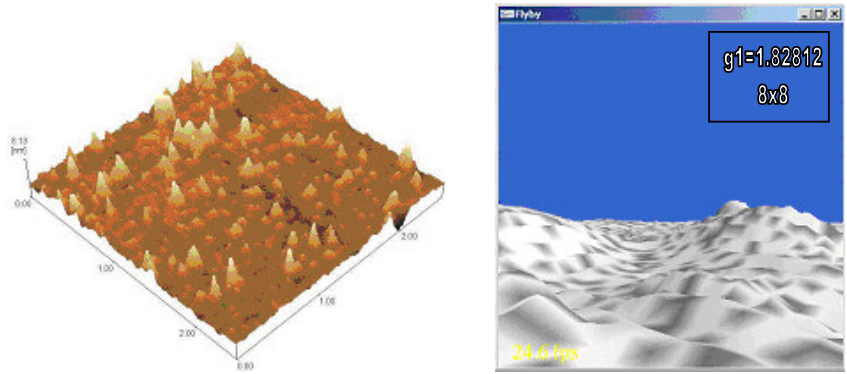

**Figure 4**: SFM structural pattern of a porous diamond-like carbon (left) and and a flyby snapshot (right), using the StrFB code, computing the level of correspondent local asymmetry in an area of 8x8.

## 5    CONCLUSIONS

This new computational tool view different angles by interactively flying through the structure, measuring its local asymmetric fragmentation. The visualization and measurement of asymmetric fragmentation can be easily included in the StrFB code for a fine investigation of structural differences among samples with very high complex porosity patterns. Among the existing techniques this seems to be the most sensible for detailed analysis of space structures in the nanometrics scales. In summary we have shown that the gradient pattern analysis technique in a flyby environment is a reliable method to investigate, qualitatively and quantitatively, the morphology of π-Si active porosity and "porous diamond-like" carbon. Taking into account the characterization of phenomena as the PSA we stress the importance to discuss the applicability of this approach into the field of nanofabrication.

**Acknowledgements** The authors acknowledges financial support from from FAPESP, CNPq, CNPq/NanoSemiMat under grant no. 550.015/01-9, the Swedish Foundation for International Cooperation in Research and Higher Education (STINT) and Swedish Research Council (VR).